\documentclass[fleqn,usenatbib]{mnras}

\usepackage[T1]{fontenc}
\usepackage{ae,aecompl}

\usepackage{graphicx}	
\usepackage{amsmath}	
\usepackage{newtxmath}
\usepackage{amssymb}	
\usepackage{booktabs}
\usepackage{tabularx}
\usepackage{layouts}
\usepackage{multirow}
\usepackage{dcolumn}
\usepackage{lscape}
\usepackage{stackengine}
\usepackage{nicefrac}
\hypersetup{pdfauthor={S. J. D. Purser},
               pdftitle={Constraining the nature of DG Tau A's thermal and non-thermal radio emission},
               pdfkeywords={stars: low-mass, stars: individual (DG Tau A), stars: jets, stars: formation, radio continuum: general, radiation mechanisms: non-thermal},
               bookmarksnumbered=true}

\stackMath

\newcolumntype{Y}{>{\raggedright\arraybackslash}X}
\newcolumntype{Z}{>{\centering\arraybackslash}X}
\newcommand{\pc}{{\rm\thinspace pc}}
\newcommand{\cm}{{\rm ~ cm}}
\newcommand{\km}{{\rm ~ km}}

\newcommand{\metre}{{\rm ~ m}}
\newcommand{\kmps}{\hbox{${ ~ \rm\km\s^{-1}}$}}
\newcommand{\cmc}{\hbox{${ ~ \rm\cm^{-3}}$}}

\newcommand{\s}{{\rm ~ s}}
\newcommand{\au}{{\rm ~ au}}

\newcommand{\yr}{{\rm ~ yr}}

\newcommand{\eV}{{\rm ~ eV}}
\newcommand{\mJy}{{\rm ~ mJy}}
\newcommand{\uJy}{{\rm ~ \upmu Jy}}

\newcommand{\Msol}{\hbox{${\rm ~ M_{\odot}}$}}
\newcommand{\Lsol}{\hbox{${\rm ~ L_{\odot}}$}}

\newcommand{\mas}{{\rm ~ mas}}

\newcommand{\rahr}{\hbox{$\rm^h$}}
\newcommand{\ramin}{\hbox{$\rm^m$}}

\newcommand{\GHz}{{\rm ~ GHz}}
\newcommand{\MHz}{{\rm ~ MHz}}

\title[New radio studies of DG Tau A's jet]{Constraining the nature of DG Tau A's thermal and non-thermal radio emission}

\author[S. J. D. Purser et al.]{S. J. D. Purser,$^{1}$\thanks{E-mail: purser@cp.dias.ie} 
R. E. Ainsworth,$^{1,2}$ 
T. P. Ray,$^{1}$
D. A. Green,$^{3}$
A. M. Taylor$^{1,4}$
\newauthor
and A. M. M. Scaife$^{2}$
\\
$^{1}$School of Cosmic Physics, Dublin Institute for Advanced Studies, 31 Fitzwilliam Place, Dublin 2, Ireland\\
$^{2}$Jodrell Bank Centre for Astrophysics, Alan Turing Building, School of Physics and Astronomy, University of Manchester,\\~~Oxford Road, Manchester, M13 9PL, United Kingdom\\
$^3$Astrophysics Group, Cavendish Laboratory, 19 J. J. Thomson Avenue, Cambridge CB3 0HE\\
$^4$DESY, Platanenallee 6, D-15738 Zeuthen, Germany
}

\date{Accepted XXX. Received YYY; in original form ZZZ}

\pubyear{2018}

\volume{{\rm in press}}
\begin{document}
\label{firstpage}
\pagerange{\pageref{firstpage}--\pageref{lastpage}}
\maketitle

\begin{abstract}
DG Tau A, a class-II young stellar object (YSO) displays both thermal, and non-thermal, radio emission associated with its bipolar jet. To investigate the nature of this emission, we present sensitive ($\upsigma\sim2\uJy \,\mathrm{beam}^{-1}$), Karl G.\ Jansky Very Large Array (VLA) $6$ and $10\GHz$ observations. Over $3.81\yr$, no proper motion is observed towards the non-thermal radio knot C, previously thought to be a bowshock. Its quasi-static nature, spatially-resolved variability and offset from the central jet axis supports a scenario whereby it is instead a stationary shock driven into the surrounding medium by the jet. Towards the internal working surface, knot A, we derive an inclination-corrected, absolute velocity of $258\pm23\kmps$. DG Tau A's receding counterjet displays a spatially-resolved increase in flux density, indicating a variable mass loss event, the first time such an event has been observed in the counterjet. For this ejection, we measure an ionised mass loss rate of $(3.7\pm1.0) \times 10^{-8} \Msol \yr^{-1}$ during the event. A contemporaneous ejection in the approaching jet isn't seen, showing it to be an asymmetric process. Finally, using radiative transfer modelling, we find that the extent of the radio emission can only be explained with the presence of shocks, and therefore reionisation, in the flow. Our modelling highlights the need to consider the relative angular size of optically thick, and thin, radio emission from a jet, to the synthesised beam, when deriving its physical conditions from its spectral index.
\end{abstract}


\section{Introduction}
\label{sec:intro}
At radio wavelengths, partially-ionised jets are almost ubiquitously \citep{Anglada1995,Furuya2003,AMI2011a} observed towards class 0, I and II protostars. Launched as by-products of accretion processes, these phenomena are highly collimated \citep[e.g.\ opening angles of $3\degr$ to $4\degr$ after initial collimation in the cases of RW Aur and CW Tau,][]{Dougados2000}, high-velocity \citep[e.g.\ HH1 and HH2,][where proper motions of $\sim200-400\kmps$ were observed]{Bally2002} outflows which carry away both material and angular momentum, aiding the accretion of material by a young stellar object (YSO). With optical and near-infrared line observations prevalent in the literature, many of their physical properties are relatively well known. Mass loss rates ($\dot{M}_\mathrm{jet}$), ionisation fractions ($\chi_{\rm i}$) and hydrogen total densities ($n_\mathrm{H}$) have been widely deduced from line ratios, with typical values calculated to be between $10^{-9}\,\Msol\yr^{-1}\leq\dot{M}_\mathrm{jet}\leq10^{-6}\,\Msol\yr^{-1}$ \citep[dependent on evolutionary class and YSO mass,][]{Caratti2012}, $0.02\leq\chi_{\rm i}\leq0.4$ \citep{Hartigan1994,Bacciotti1999} and $10^3\cmc\leq n_\mathrm{H} \leq 10^5\cmc$ \citep{Bacciotti1999} respectively.

As for the exact mechanism(s) responsible for the jets' launch and collimation, there is still much uncertainty. Radio observations of a high-mass YSO have revealed a large-scale, poloidal, magnetic field at large ($10^4-10^5\au$) scales \citep[HH 80--81,][]{CarrascoGonzalez2010}, the field lines of which were aligned along the outflow axis and possessed field strengths on the order of $\sim0.2\,\mathrm{mG}$. This configuration is in agreement with either a disc-wind \citep{BlandfordPayne1982,Pudritz1983} or X-wind \citep{Shu1994} launching/collimation model, whereby ionised material is magneto-centrifugally accelerated along magnetic field lines rooted in the disc and/or protostar. Considering this single example of this type of observation, further detections of such field configurations, especially towards low-mass YSOs, are of paramount importance in constraining jet models.

In this work we study the class-\textsc{II} classical T-Tauri star, DG Tau A, which is located at the eastern tip of the L1495 filament of the Taurus Molecular Cloud (TMC). Previously, distance estimates of $140-150\pc$ were adopted, however, with Gaia \citep{GaiaCollaboration2016,GaiaCollaboration2018b} this estimate was refined to $120.8\substack{+2.2 \\ -2.1}\ \pc$ by \citet{BailerJones2018}, who correctly handled the asymmetric probability distributions present in Gaia DR1/DR2. With this new distance, DG Tau A's luminosity is further refined to possesses a bolometric luminosity of $4.7\Lsol$ \citep[from $6.4\Lsol$ calculated by][who used a distance of $140\pc$]{KenyonHartmann1995}. It is known to harbour a clockwise-rotating (from the observer's perspective) jet, the outflow axis of which lies along a position angle of $223\degr$ according to \citet{Bacciotti2002} and at an inclination of $37.7\pm2\fdg2$ \citep{EisloffelMundt1998}. Velocity and density gradients are present across the approaching (southwest) jet's cross-section whereby a high-velocity component ($v\simeq220\kmps\!,\,\,n\simeq10^6\cmc$) is enveloped within a lower-velocity component ($v\simeq100\kmps\!,\,\,n\simeq5\times10^5\cmc$) observed, in the optical, using the Hubble Space Telescope \citep[HST;][who also inferred a total jet mass loss rate of $1.3\times10^{-7}\Msol\yr^{-1}$]{Coffey2008}. X-ray observations revealed both a hard and soft component \citep{Gudel2005} to the spectrum with the former originating in a magnetically confined corona above the star and latter thought to result from shocks at the jet base \citep[later confirmed by][who observed a separation of $\sim50\au$ between the hard and soft X-ray components]{SchneiderSchmitt2008}. Later Chandra X-ray images showed a bipolar X-ray jet extending out to a distance of $5\arcsec$ either side of the star, with the receding, northeast counter-jet being weaker and spectrally harder as a result of absorption by a dust disc \citep{Gudel2008}. Subsequently, this dust disc was directly observed by the Combined Array for Research in Millimeter-wave Astronomy (CARMA) at $230\GHz$ and derived to have a major-axis position angle, radius and inclination (angle between the disc's rotation axis and line of sight) of $\sim120\degr$, $\sim70\au$ and $\sim30\degr$ respectively \citep{Isella2010}.

\begin{figure}
\includegraphics[width=\columnwidth]{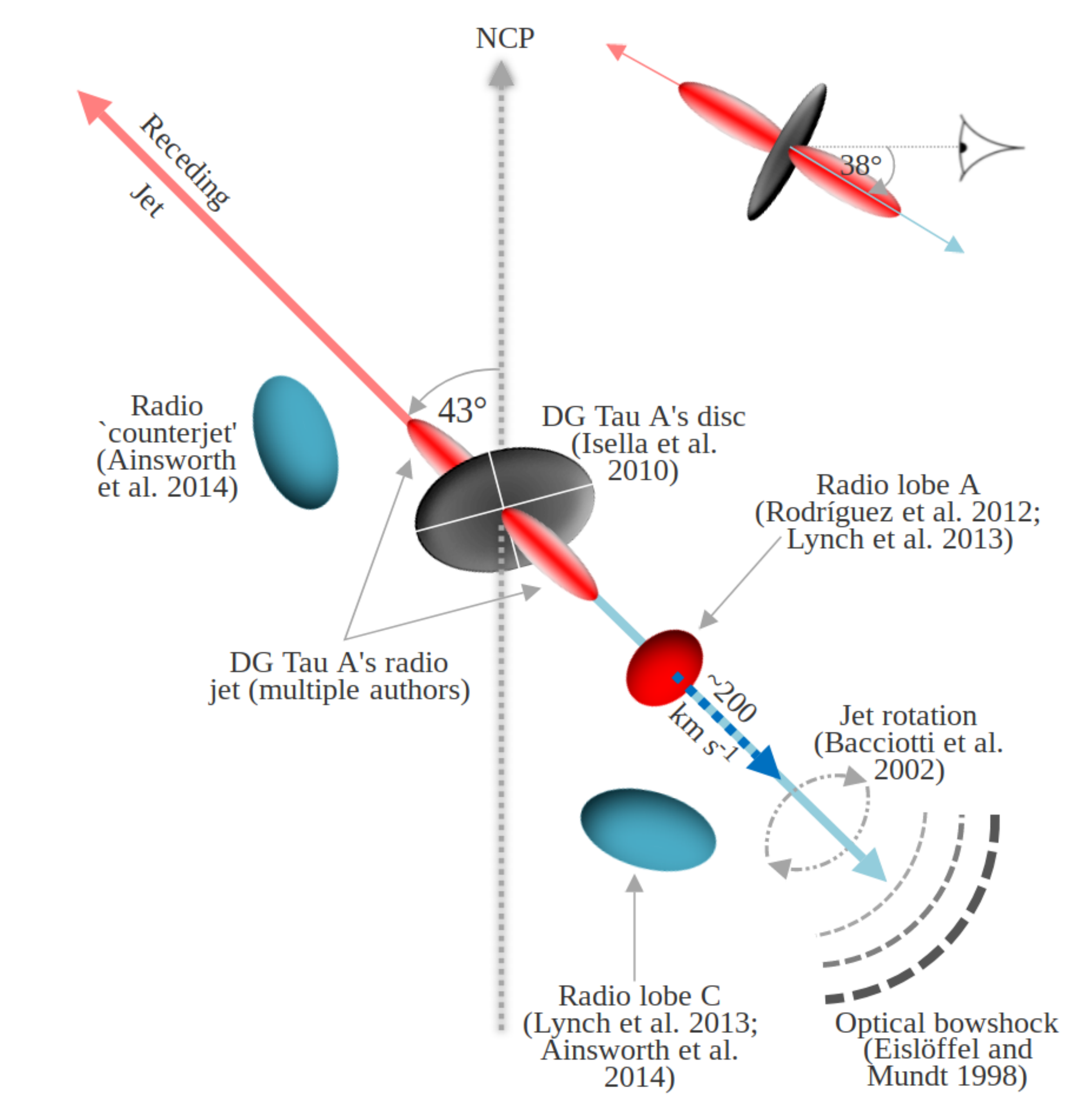}
\caption{A simplified schematic of DG Tau A's disc/jet from the observer's perspective. Inclination of the system is illustrated by the side profile in the top right corner.}
\label{fig:dgtauschematic}
\end{figure}

Spatially distinct from the thermal jet at DG Tau A's position are multiple knots of emission seen at a variety of wavelengths. H$\alpha$ observations by \citet{MundtFried1983} revealed a Herbig--Haro (HH) knot located $8\arcsec$ from DG Tau A at a position angle of $228\pm2\degr$. This knot appeared to be connected by a `light bridge' to DG Tau A, which was determined to be part of the jet's stream (one of the first examples of such) and named HH 158. Using spectroscopic observations, the same work showed that the emitting gas close to DG Tau A was moving with a velocity of $-250\pm10\kmps$, and this was subsequently found to be typical of jets from low-mass YSOs. \citet{Rodriguez2012DGTauA} observed $[\ion{S}{II}]$ $6716$\AA$\,$ and $6731$\AA$\,$ emission lines, detecting two knots separated from DG Tau A by $\sim7\arcsec$ \citep[knot k0, first observed by][]{SolfBohm1993} and $13\arcsec$ \citep[the knot initially reported by][]{MundtFried1983} along the previously established jet position angle. The same authors compiled positional data from optical, radio and X-ray observations over a period of $\sim20\yr$ and showed that the southwest knot, separated from DG Tau A by $7\arcsec$, was moving with a proper motion of $159\pm7\kmps$ (adapted for a distance of $120.8\pm2.2\pc$, rather than the $150\pc$ adopted in their paper).

In the radio regime DG Tau A has been the target of many observing programmes, the first being that of \citet{Cohen1982} who detected an elongated radio source centred on DG Tau A ($S_\mathrm{4.9GHz} \sim 0.7\mJy$). Later, radio observations derived a thermal, spectral index of $\alpha\sim0.5$ \citep{CohenBieging1986,Lynch2013} for the radio source, typical of \citep{Reynolds1986}, and showing it to be, an ionised jet. Multi-epoch studies \citep{Rodriguez2012DGTauA} showed the jet's radio emission to be highly variable over $30\yr$ and modelling suggested this may be the result of a sinusoidal variation in ejection velocity, leading to flux density variations with a period of $4.8\pm0.3\yr$. That work's findings were in agreement with the results of a preceding study by \citet{Raga2001} who invoked a precessing, and variable velocity, jet model. Synthetic maps, in H$\upalpha$ and [O\textsc{i}] $\uplambda 6300$, of that model reproduced many of the observations of \citet{Dougados2000} and supported the idea of reionisation of material along the jet's axis. \citet{Lynch2013} revealed a spatially distinct knot, C, separated from the thermal radio jet by $14\arcsec$ to the southwest but displaced from the jet axis. A closer knot of radio emission \citep[knot A, coinciding with knot k0 from][]{Rodriguez2012DGTauA} $7\arcsec$ to the southwest was also detected. Their work established linear polarization limits on the thermal jet of $<2$ percent and $\lesssim50$ percent for knot C (estimated from their clean maps). Giant Metrewave Radio Telescope (GMRT) observations \citep{Ainsworth2014} showed the previously detected knot C to be non-thermal ($\alpha=-0.9\pm0.1$) and concluded it to be a bow shock on account of its curved morphology and proximity to the extrapolated position of the prominent bow shock reported by \citet{EisloffelMundt1998}. Further analysis \citep[by][]{Ainsworth2014} calculated that the shock was enough to produce a significant flux of low energy cosmic rays which, when extrapolated over the Galactic star formation rate and average molecular cloud lifetimes, provided a local (i.e.\ within the cloud) energy density close to that of the ISM ($\sim10^{-2}\eV\cmc$). If this interpretation is correct, it would provide a significant source of low energy cosmic rays, other than supernovae. \autoref{fig:dgtauschematic} summarises the most relevant previous observations of DG Tau A as a schematic illustration.

In light of previous radio works targeting DG Tau A, this paper therefore aims to answer the following questions. Can we confirm the bowshock nature of knot C through the detection of proper motions? Is the shocked knot, knot A, the result of periodic changes in outflow velocity, or is it simply an evolving shock? Analogous to a high-mass example (HH 80--81), can we detect magnetic field directions and strengths at the shock sites? Are there fainter shock-sites present which have previously remained undetected? In \autoref{sec:obs} we explain the observational setup of the Karl G.\ Jansky Very Large Array (VLA) observations, for which our results are shown explicitly in \autoref{sec:results}. Following this is a discussion (\autoref{sec:dicussion}) of the implications of these results upon the questions posed above and we finish with our conclusions in \autoref{sec:conclusions}.

\section{Observations and Data Reduction}
\label{sec:obs}
All observations were conducted using the VLA under project ID 16A-051, while in its C-configuration, with baseline lengths between $35$ and $3400\metre$. Two frequency setups in full polarization mode ($\mathrm{LL}$, $\mathrm{RR}$, $\mathrm{RL}$ and $\mathrm{LR}$) were utilised, one with a central frequency of $6\GHz$, and the other with a central frequency of $10\GHz$. Both the $6$ and $10\GHz$ setups employed bandwidths of $4\GHz$ divided into 32 sub-bands of $64\times2\MHz$ channels each. All $10\GHz$ observations were conducted in a single run on 2016 February $6-7$ (epoch 2016.10), while the $6\GHz$ observations were conducted in two separate runs on the 2016 February 21 and 2016 February $25-26$ (epoch 2016.15). Total integration times were $178.1\min$ and $86.1\min$ for the $6$ and $10\GHz$ observations respectively.

Our observing strategy was to observe the flux-density/bandpass calibrator, $3\mathrm{C}147$, at both the beginning and end of each set of observations. Complex gains were calibrated using observations of $\mathrm{J}0403+2600$ (angular separation of $5.4$ degrees from DG Tau A) which was observed every $25\min$ (2016 February 21) or $35\min$ (2016 February $25-26$) at $6\GHz$, and $30\min$ at $10\GHz$. Our $6\GHz$ observations also used calibrators for instrumental polarization/leakage ($3\mathrm{C}147$) and absolute polarization angle ($3\mathrm{C}138$). Observing information for all calibrators and DG Tau A is shown in \autoref{tab:obsfields}.

\begin{table}
\centering
\caption{A table of all observed fields of our $6$ and $10\GHz$ observations, with their designations (column 1), positions (columns 2 and 3) and observational purposes (column 4).}
\begin{tabular}{cccc}
\toprule
Name & R.A. & Dec. & Calibrator or \\
 & $\mathrm{\left[J2000\right]}$ & $\mathrm{\left[J2000\right]}$ & Target Type \\
\midrule
\multirow{3}{*}{$3\mathrm{C}147$} & \multirow{3}{*}{5\rahr42\ramin36\fs138} & \multirow{3}{*}{+49\degr51\arcmin07$\farcs$23} & Flux-density \\
 & & & Bandpass \\
 & & & Pol. leakage \\
\multirow{1}{*}{$3\mathrm{C}138$} & \multirow{1}{*}{5\rahr21\ramin09\fs886} & \multirow{1}{*}{+16\degr38\arcmin22$\farcs$05} & Abs.\ pol.\ angle\\
J0403$+2600$ & \multirow{1}{*}{4\rahr03\ramin05\fs586} & \multirow{1}{*}{+26\degr00\arcmin01$\farcs$50} & \multirow{1}{*}{Complex gain}\\
DG Tau A & 4\rahr27\ramin04\fs693 & +26\degr06\arcmin15$\farcs$82 & Science target \\
\bottomrule
\end{tabular}
\label{tab:obsfields}
\end{table}

\begin{figure*}
	\includegraphics[width=5.329in]{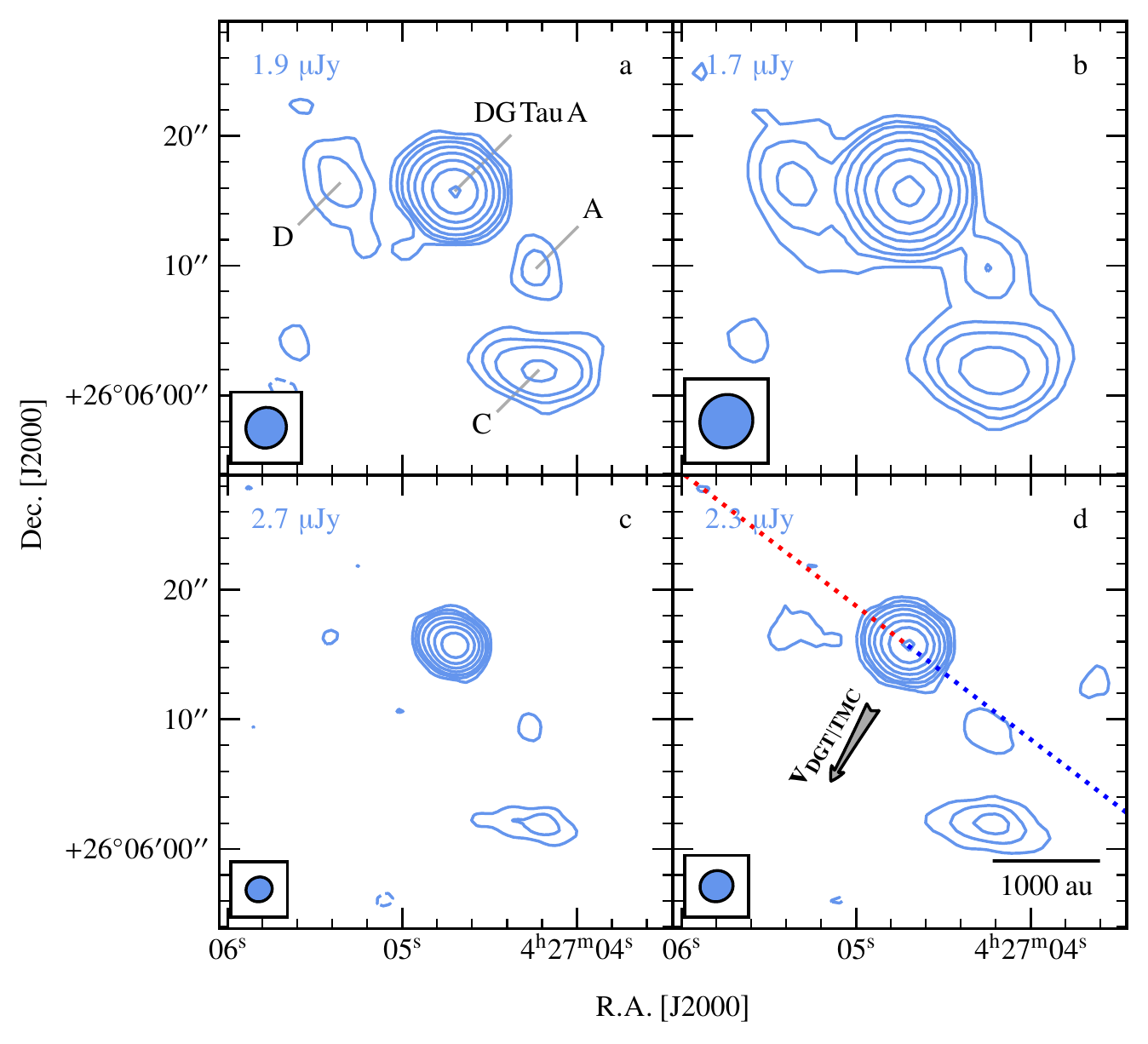}
	\caption{Contour plots of flux density at $6\GHz$ (panels a and b) and at $10\GHz$ (c and d). Two values for the robust parameter were imaged of 0.5 (panels a and c) and 2 (panels b and d). Restoring beams used were $3\farcs20\times3\farcs07$ at a position angle of $-42\fdg\;\!\!2$, $4\farcs18\times4\farcs01$ at a position angle of $-40\fdg\;\!\!1$, $2\farcs05\times1\farcs89$ at a position angle of $-65\fdg\;\!\!9$ and $2\farcs57\times3\farcs39$ at a position angle of $-69\fdg\;\!\!1$ for panels a, b, c and d respectively. Image noise levels ($\upsigma_\mathrm{RMS}$) are displayed in the top left corner of each plot and contours are set at $-3, 3, 6, 10, 20, 40, 80, 160$ and $320\times\upsigma_\mathrm{RMS}$. In panel d, the motion (see \autoref{sec:nonthermal}) of DG Tau A relative to the TMC is indicated by a grey arrow and the red/blue dotted lines show the deconvolved position angle (i.e.\ current jet axis) for DG Tau A's receding/approaching jet.}
	\label{fig:cplots2016}
\end{figure*}

For the data reduction process, the Common Astronomy Software Applications (\textsc{casa}) package \citep{CASARef} was utilised throughout. Obviously bad visibilities were initially flagged manually, supplemented with the use of \textsc{casa}'s \texttt{tfcrop} algorithm and, after an initial round of calibration, then reflagged both manually and using \textsc{casa}'s \texttt{rflag} algorithm. After a second round of calibration, solutions were inspected and verified to smoothly vary with time and/or frequency, and then applied to the data. Corrected visibilities were subsequently split off into a separate, calibrated measurement set. For both the 6 and $10\GHz$ data, the DG Tau A field was imaged out to the primary beam Full-Width at Half Maximum (FWHM; $420\arcsec$ and $252\arcsec$, at the central frequencies, respectively), and field sources with high signal-to-noise ratios were cleaned from the dirty images to form a self-calibration model which was subsequently used to calculate initial, phase-only, self-calibration solutions with one solution per target scan. After inspecting the quality of these new calibration tables, they were applied to the data, a new measurement set was created from the self-calibrated data, the DG Tau A field was re-imaged, and the self-calibration process was reiterated for a further round of phase-only, and two rounds of phase and amplitude, self-calibration. For both frequencies, the calibration solutions converged at this point.

Synthesised beam widths were typically $3\farcs1$ and $1\farcs9$ and the maximum recoverable angular scales are $240\arcsec$ and $145\arcsec$ for 6 and $10\GHz$ respectively. We therefore do not expect flux loss, as a result of incomplete $uv$-sampling, to be a significant issue for the typical spatial scales of DG Tau A's radio emission. 

Many of the proceeding analyses are direct comparisons between archival and our observations. For these juxtapositions of knot positions, flux densities and dimensions, we used the archival data of \citet{Lynch2013} (project ID TDEM0016) which was observed at central frequencies of $5.5\GHz$ (epoch 2012.22, $\upDelta t=3.93\yr$) and $8.5\GHz$ (epoch 2012.29, $\upDelta t=3.81\yr$). We downloaded, reduced, self-calibrated (as per the method described above) and then re-imaged, at both a robustness ($R$) of $-2$ and $2$, these data. Since the VLA was not fully upgraded during the 2012A semester, the frequency coverage of those observations ($2\GHz$ bandwidth) was only half of that of our 2016A data. Therefore to compare the two datasets, a subset of the 2016 data was re-imaged using identical frequency coverages as the 2012 data in order to closely mirror the $uv$-sampling of the older epoch. As a note, the 2012 X-band data had low amplitudes recorded for half the bandpass ($9\GHz-10\GHz$) which was unsalvageable and therefore completely flagged. From this point on, these comparative data are referred to as $5.5\GHz$ and $8.5\GHz$ datasets.

\section{Results}
\label{sec:results}
\begin{table*}
\caption{A table of the \textsc{imfit}-derived positions (columns 2 and 4), positional errors (columns 3 and 5), peak fluxes (column 6), integrated fluxes (column 7) and dimensions (columns 8--10) for DG Tau, and its associated lobes of emission, for our 2016 data at both observing bands. These quantities were derived from the clean images utilising a robustness of $0.5$. Errors do not include the uncertainty in the absolute flux scaling.}
\begin{center}
\begin{tabular}{l c c c c c D{,}{\,\pm\,}{-1} D{,}{\,\pm\,}{-1} D{,}{\,\pm\,}{-1} D{,}{\,\pm\,}{-1} c}
\toprule
Component & R.A.\ & $\upDelta$\ R.A.\ & Dec. & $\upDelta$\ Dec. & $S^\mathrm{peak}_\upnu$ & \multicolumn{1}{c}{$S^\mathrm{int}_\upnu$} & \multicolumn{1}{c}{$\theta_\mathrm{maj}$} & \multicolumn{1}{c}{$\theta_\mathrm{min}$} & \multicolumn{1}{c}{$\theta_\mathrm{PA}$}\\
 & $\mathrm{\left[J2000\right]}$ & $\left[\mathrm{mas}\right]$ & $\mathrm{\left[J2000\right]}$ & $\left[\mathrm{mas}\right]$ & $\left[\!\!\uJy\right]$ & \multicolumn{1}{c}{$\left[\!\!\uJy\right]$} & \multicolumn{1}{c}{$\left[\arcsec\right]$} & \multicolumn{1}{c}{$\left[\arcsec\right]$} & \multicolumn{1}{c}{$\left[\degr\right]$}\\
\midrule
\multicolumn{10}{c}{\textit{C--band}}\\
\midrule
DG Tau A & 04\rahr 27\ramin 04\fs6998 & 5 & +26\degr 06\arcmin 15\farcs739 & 4 & 658 & 748,4 & 1.63,0.03 & 0.40,0.10 & 53,2\\
A & 04\rahr 27\ramin 04\fs2350 & 153 & +26\degr 06\arcmin 09\farcs771 & 251 & 16 & 21,4 & <3.50 & <1.50 & -\\
C & 04\rahr 27\ramin 04\fs2162 & 148 & +26\degr 06\arcmin 01\farcs973 & 60 & 46 & 95,6 & 5.30,0.41 & 1.32,0.42 & 81,3\\
D & 04\rahr 27\ramin 05\fs3541 & 116 & +26\degr 06\arcmin 16\farcs408 & 182 & 16 & 42,4 & 5.34,0.53 & 2.59,0.41 & 22,6\\
\midrule
\multicolumn{10}{c}{\textit{X--band}}\\
\midrule
DG Tau A & 04\rahr 27\ramin 04\fs7006 & 5 & +26\degr 06\arcmin 15\farcs740 & 4 & 774 & 932,8 & 1.27,0.03 & 0.34,0.09 & 53,2\\
A & 04\rahr 27\ramin 04\fs2690 & 175 & +26\degr 06\arcmin 09\farcs371 & 242 & 12 & 20,6 & 2.17,0.72 & 0.90,0.77 & 21,27\\
C & 04\rahr 27\ramin 04\fs2470 & 497 & +26\degr 06\arcmin 01\farcs879 & 142 & 24 & 72,15 & 5.37,1.29 & 1.20,0.87 & 82,6\\
D & 04\rahr 27\ramin 05\fs4170 & 449 & +26\degr 06\arcmin 16\farcs172 & 573 & 10 & 24,9 & 4.15,1.96 & 1.16,0.87 & 145,26\\
\bottomrule
\end{tabular}
\end{center}
\label{tab:results2016}
\end{table*}

\begin{figure*}
	\includegraphics[width=5.329in]{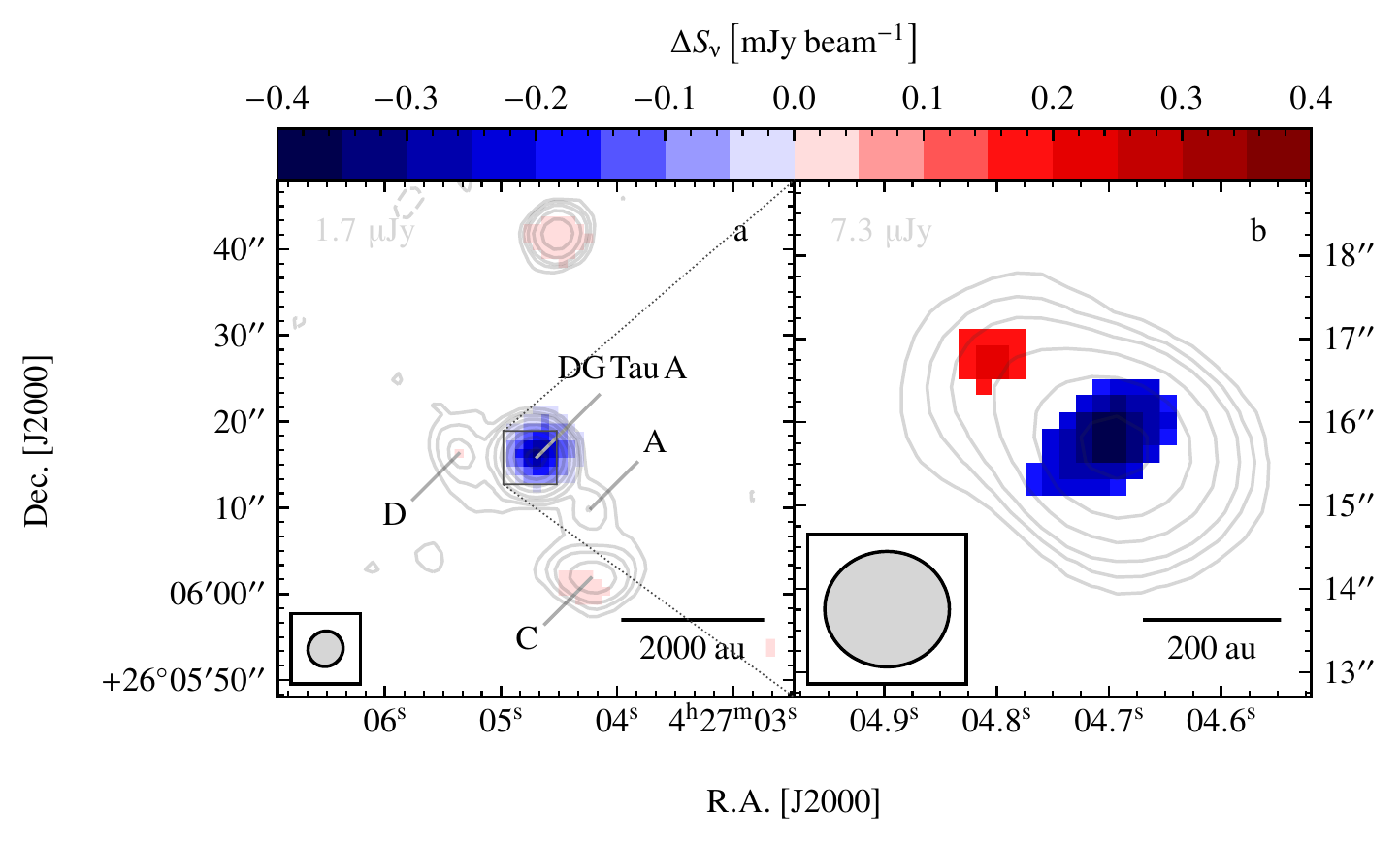}
	\caption{A map of the $>3\upsigma$ flux density differences between $2012$ and $2016$ flux density maps (colourscale). For this analysis, identical bandwidths were utilised. \textit{Panel a}: Differences between the two epochs for the $5.5\GHz$ data, utilising $R=2$. Contours show the $6\GHz$ (full-bandwidth) 2016 C-band image with levels and restoring beam sizes as in panel b of \autoref{fig:cplots2016}; \textit{Panel b}: Differences between the two epochs for $8.5\GHz$ data, utilising $R=-2$. Contours show the $10\GHz$ (full-bandwidth), uniformly-weighted image from 2016 set at $-3, 3, 5, 7, 10, 20, 40$ and $80\upsigma$ where $\upsigma=7.3\uJy\,\mathrm{beam^{-1}}$. The restoring beam used was $1\farcs495 \times 1\farcs384$ at $\theta_\mathrm{PA} = 89\fdg\;\!\!7$.}
	\label{fig:allbandsfluxdiff}
\end{figure*}

After the calibration procedures outlined in \autoref{sec:obs}, the $6$ and $10\GHz$ 2016 data were imaged using \textsc{casa}'s \texttt{clean} task with $R=0.5$ and $R=2$, with the former being a compromise between spatial resolution/sensitivity and the latter prioritising sensitivity and the recovery of emission from larger angular scales. Resulting clean images\footnote{All images present in this work are available at doi:\url{10.5281/zenodo.1321756}} (\autoref{fig:cplots2016}) show 4 distinct components, DG Tau A, knot A, knot C and knot D, detected in the vicinity (within $15\arcsec$) of the pointing centre. Names are taken from the literature, apart from knot D which is identified as the `counterjet' detected, using GMRT observations at $325\MHz$, by \citet{Ainsworth2014}. All components' positions, flux densities and dimensions were measured using the \texttt{imfit} task of \textsc{casa}, the results of which are presented in \autoref{tab:results2016}. \texttt{imfit} works by fitting a Gaussian to components in the image plane, subsequently deconvolving that fitted Gaussian from the beam to estimate the dimensions of the emission \citep[error estimation is based on the work by][]{Condon1997}. The indices $\alpha$ and $\gamma$, defined by $S_\upnu\propto\nu^\alpha$ and $\theta_\mathrm{maj}\propto\nu^\gamma$ (where $\theta_\mathrm{maj}$ is the deconvolved major axis), were also computed and are presented in \autoref{tab:spix}. We have also calculated $\alpha$ for knots C and D using the GMRT results of \citet{Ainsworth2014} at $325\MHz$ for knot D and at both $325$ and $610\MHz$ for knot C. 

For comparison of the $5.5$ and $8.5\GHz$ datasets from the 2012 and 2016 epochs, positions, flux densities and deconvolved dimensions of DG Tau A, knot A and knot C were measured using \texttt{imfit}, the (naturally-weighted) results for which are shown in Table 1 of online-only supporting material.

Stokes Q, U and V images were also made from the $6\GHz$ data, however no polarization, linear or circular, was detected towards any component with $3\upsigma$ upper limits on the linear polarization fraction of $<1.3$, $<50.8$, $<18.2$ and $<51.5$~percent for DG Tau A, A, C and D respectively. For circular polarization these limits are $<0.9$, $<35.4$, $<12.7$ and $<35.9$~percent respectively.

\begin{table}
	\centering
	\caption{A table of the derived (via the method of least squares) indices values of all sources associated to DG Tau A for flux density ($\alpha$, column 2) and major axis length ($\gamma$, column 3), between $6$ and $10\GHz$.}
	\begin{minipage}{\columnwidth}
	\renewcommand*\footnoterule{}
	\begin{tabularx}{\columnwidth}{c D{,}{\,\pm\,}{-1} D{,}{\,\pm\,}{-1}}
		\toprule
		\multicolumn{1}{X}{\centering Component}  &  \multicolumn{1}{X}{\centering $\alpha$} & \multicolumn{1}{X}{\centering $\gamma$} \\
		\midrule
		DG Tau A & +0.43,0.16 & -0.49,0.05 \\
		A      & -0.10,0.83 & >-0.94     \\ 
		C$^\dagger$      & -0.54,0.43 & +0.03,0.49 \\
		D$^\dagger$      & -1.10,0.76 & -0.49,0.94 \\
		\bottomrule
	\end{tabularx}
	\end{minipage}
	\scriptsize{$^\dagger$When including GMRT data from \citet{Ainsworth2014}, values for $\alpha$ of $-0.89\pm0.08$ and $-0.91\pm0.11$ are calculated for knots C and D respectively.}
	\label{tab:spix}
\end{table}

\section{Discussion}
\label{sec:dicussion}
\subsection{Flux Density/Morphology Variability}
\label{sec:variability}
DG Tau A has been previously established to possess a flux-variable, radio jet (see \autoref{sec:intro}). Flux-variability was proposed to be the result of periodic ejections (every $\sim5\yr$) of material in the approaching (southwest) jet, for which several, optical knots of emission have been identified by multiple authors \citep[see][specifically their Table 2 and Figure 5]{Rodriguez2012DGTauA}. From our results (see our supplementary, online-only, Table 1), we observe a decrease in DG Tau A's \texttt{imfit}-derived, integrated flux density at both $5.5$ and $8.5\GHz$ of $-325\pm63\uJy$ and $-327\pm76\uJy $ respectively, from the $R=2$ images. Otherwise, no significant ($>3\upsigma$) variability is observed apart from a $\sim2\upsigma$ integrated flux density increase for knot C of $+42\pm21 \uJy$ at $5.5\GHz$ and $>+59\pm21\uJy$ at $8.5\GHz$. However knot C is significantly extended and therefore this \texttt{imfit}-derived quantity may be less reliable if its true morphology significantly deviates from a Gaussian. An alternative approach to deal with non-Gaussian morphology is to integrate the flux in a region encapsulating the $3\upsigma$ emission from the 2012 epoch. This shows a similar flux increase from 2012 to 2016 of $+28\pm13\uJy$ at $5.5\GHz$ and $>+30\pm12\uJy$ at $8.5\GHz$. As a note, all errors in the changes in integrated flux density include a $5$~percent uncertainty in the absolute flux density scale. Through standard propagation of errors (measurements in each epoch are independent), this uncertainty is incorporated into the relevant analyses/errors presented throughout this work.

\begin{figure*}
	\includegraphics[width=5.329in]{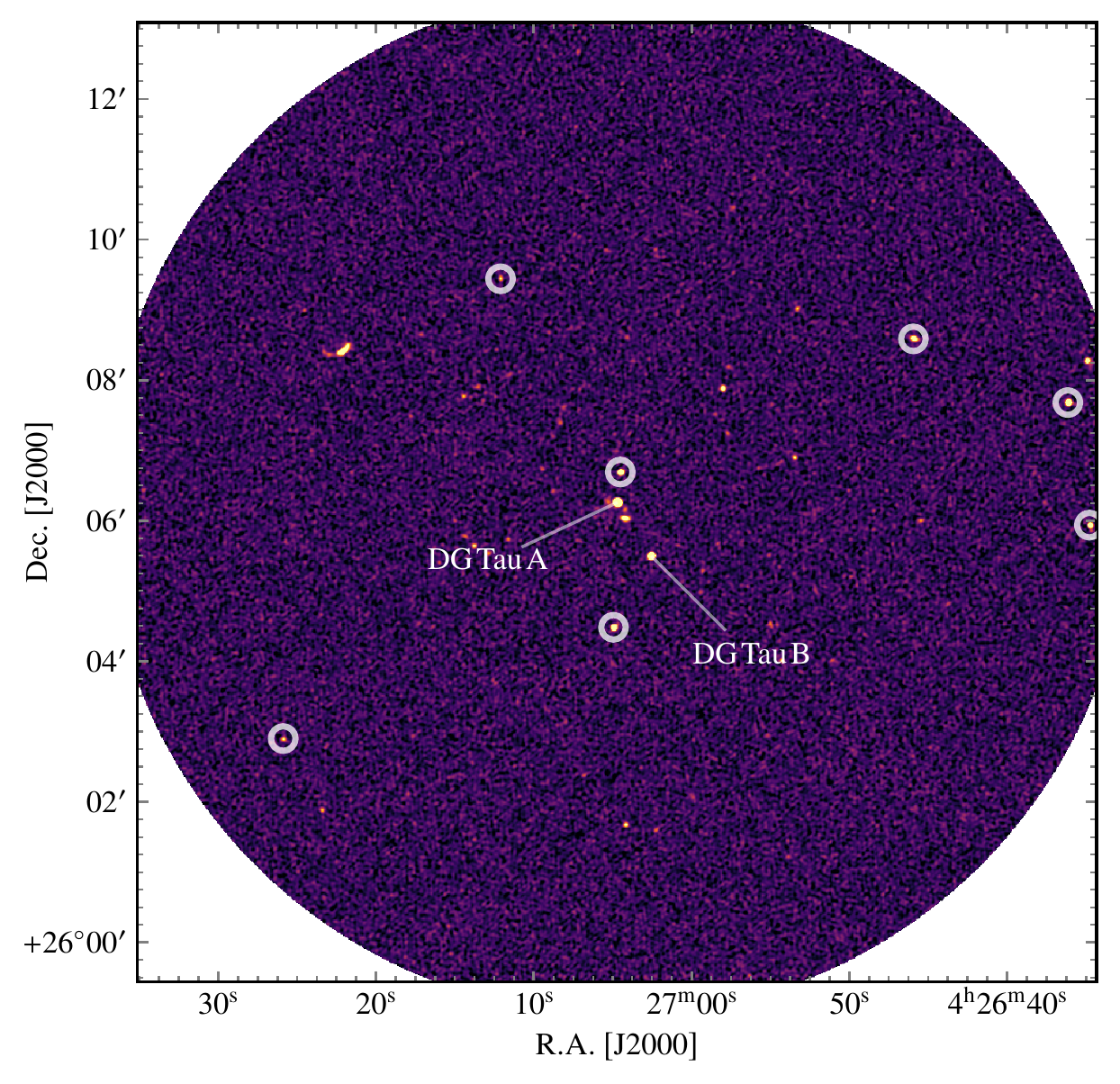}
	\caption{$5.5\GHz$ image of the field out to the $5$~percent level of the primary beam. While DG Tau A and DG Tau B are explicitly labelled, all point-like background sources used to calibrate astrometric positions between epochs are highlighted with circular markers.}
	\label{fig:cbandprimarybeam}
\end{figure*}

In order to examine the morphology of any variability between 2012 and 2016, we have created pixel-to-pixel flux density difference maps for both $5.5$ and $8.5\GHz$ data (as shown in panels a and b of \autoref{fig:allbandsfluxdiff} respectively). In order to probe variability on different spatial scales, the $5.5\GHz$ flux density difference map (panel a) used a natural robustness of $R=2$, while the $8.5\GHz$ data used a more uniform robustness of $R=-2$ (panel b). From panel a it is obvious that DG Tau A has decreased in flux density between the observations, while knot C has slightly (with respect to DG Tau A) increased in flux density ($\Delta S_\mathrm{5.5GHz}=42\pm21\uJy$). This flux density increase is asymmetric, being on the southern side of the limb-brightened `bowshock' described by \citet{Ainsworth2014}, supporting the case whereby DG Tau A's jet is impinging upon a density gradient in that direction. A point-like, unknown source $\sim25\arcsec$ to the north also shows an increase in flux density ($\Delta S_\mathrm{5.5GHz}=50\pm12\uJy$). As to the nature of this source, we derive a non-thermal spectral index for it of $\alpha=-0.44\pm0.15$, which likely means that it is extragalactic in nature, especially since it displays no proper motions (see \autoref{sec:propermotions} where it is used to calibrate positional uncertainties).

Looking at panel b where the higher resolution of the $R=-2$, $8.5\GHz$ image allows for a finer spatial analysis, the flux density variability of DG Tau A is, in fact, resolved into both a positive and negative component. The positive component is smaller in magnitude ($\delta S_\upnu^\mathrm{peak}=+211\pm62\uJy\,\mathrm{beam}^{-1}$), centred on the elongation of $8.5\GHz$ data from the 2016 epoch and separated from the negative component (presumably coincident with the YSO) by $1.78 \pm 0\farcs20$ at a position angle of $49.3\pm7\fdg\;\!\!0$. Though impossible to determine an ejection date, if the knot was ejected at an epoch of 2012.29 (simply the date of the first set of observations), a velocity in the sky's plane of $268\pm30\kmps$ is inferred. This suggests that, for the first time in DG Tau A's radio observing history, a variable ejection of jet material in the receding jet has been seen.

Assuming the ejection is optically thin (i.e.\ $\alpha=-0.1$), spherical and has a diameter of $0.84\pm0\farcs29$ ($D_\mathrm{knot}=101\pm34\au$), we calculate an emission measure of $\left(1.4\pm1.0\right) \times 10^5 \,\pc \cm^{-6}$, average electron density of $\left(2.0\pm1.1\right) \times 10^4 \,\cm^{-3}$ \citep[using Equations $1.37$ and $10.32$ to $10.34$ of][]{RohlfsWilson2012} and ionised mass of $\left(4.0\pm2.2\right)\times10^{-8}\Msol$. For these calculations we have adopted an inclination of $37.7\pm2\fdg2$ \citep{EisloffelMundt1998}, which is used throughout the rest of this work. If the ejection event took place over $1.1\pm0.4\yr$ (i.e.\ $t=D_\mathrm{knot}/v_\mathrm{jet}$), an average, ionised mass loss rate in the receding jet, during the outburst, of $\left(3.7\pm1.0\right)\times10^{-8}\Msol\yr^{-1}$ is calculated. Compared to the results of \citet{Ainsworth2013} who measured the steady, ionised mass loss rate in the approaching jet, this is only a factor of $\sim2$ greater than that estimate ($1.5\times10^{-8}\ \mathrm{M_\odot\,yr^{-1}}$). Since this is a steady-state mass-loss rate in the approaching jet, while our calculated mass-loss rate is that during an outburst event, this comparison highlights the asymmetric nature of mass loss in DG Tau A's opposing jets.


\subsection{Proper Motions}
\label{sec:propermotions}

In order to deduce accurate proper motions, errors in absolute astrometry between the two epochs (2012.22 and 2016.15) had to be compensated for. Due to the low noise levels present, 11 background sources (including DG Tau B) were detected at a $>5\upsigma$ level across the primary beam in both epochs. Positions and deconvolved sizes of these background sources were subsequently calculated, using \texttt{imfit}, for each epoch. Utilising only point-like background sources (7 of 11), the positions of which are shown in \autoref{fig:cbandprimarybeam}, overall positional changes between the two epochs were derived for each source. A weighted average of these positional changes in both right ascension and declination was then calculated. In the case whereby one of the object's shifts is the result of real proper motions, the large sample size should negate its effect upon the weighted average, since we expect the vast majority of these objects to be extragalactic in nature. From this method we therefore calculate a weighted average for the positional shift in right ascension of $0\farcs00 \pm 0\farcs05$ and in declination of $-0\farcs16 \pm 0\farcs04$. For all subsequent positional comparisons, this calculated shift is the astrometric correction applied beforehand.

\begin{table}
	\caption{A table of the position-calibrated, changes in right ascension (2\textsuperscript{nd} column), changes in declination (3\textsuperscript{rd} column), proper motion magnitudes (4\textsuperscript{th} column) and proper motion directions (5\textsuperscript{th} column) of sources associated to DG Tau A, between epochs 2012.22 and 2016.10. Naturally ($R=2$) weighted, $5.5\GHz$ images were used in these calculations.}
	\begin{tabular}{>{\centering\arraybackslash}p{1.1cm}>{\centering\arraybackslash}p{1.54cm}>{\centering\arraybackslash}p{1.54cm}>{\centering\arraybackslash}p{1.14cm}>{\centering\arraybackslash}p{1.1cm}}
		\toprule
		\footnotesize{Source}  & $\upDelta$\ R.A. &  $\upDelta$\ Dec.   & $v$             & $\theta$      \\
		        & $[\,\arcsec\,]$ &  $[\,\arcsec\,]$  & $[\!\!\!\kmps]$ & $[\,\degr\,]$ \\
		\midrule
		\footnotesize{DG$\,$Tau$\,$A} & $+0.17\pm0.05$  & $+0.01\pm0.04$  &   $25\pm7\:\;$   & $\:\;88\pm14$  \\
		A        & $-1.50\pm0.63$  & $-0.61\pm0.72$  &   $237\pm94$    &   $248\pm25$   \\
		C        & $+0.64\pm0.55$  & $-0.10\pm0.18$  &   $94\pm79\:\;$ &   $99\pm17$   \\
		\bottomrule
	\end{tabular}
	\label{tab:propermotions}
\end{table}

\begin{figure}
	\includegraphics[width=\columnwidth]{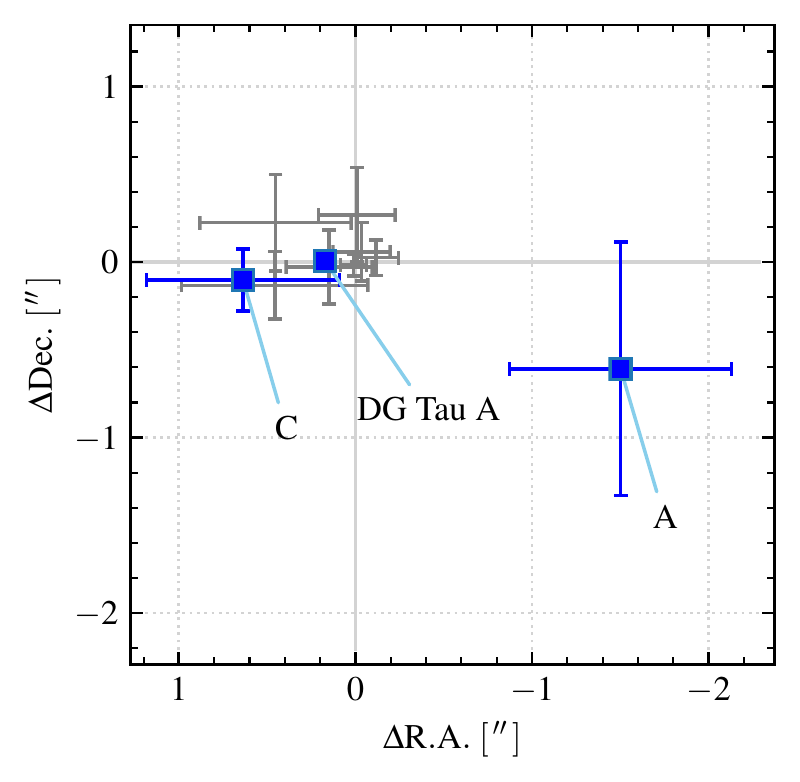}
	\caption{A plot of the position-calibrated changes in right ascension and declination of background sources (grey errorbars) and of DG Tau A, knot A and knot C (blue markers and errorbars).}
	\label{fig:propermotions}
\end{figure}

In \autoref{tab:propermotions} and \autoref{fig:propermotions} the position shifts for those sources associated with DG Tau A, which have had the astrometric correction subtracted, are shown. Using the Haversine formula for the calculation of proper motions, and adopting a distance to DG Tau A of $120.8\pm2.2\pc$ (adopted throughout this work), we deduce the velocities (in the plane of the sky) shown in the final two columns of \autoref{tab:propermotions}. It can be seen that knot A shows significant velocities between 2012.22 and 2016.10 of $237\pm94\kmps$, at a position angle of $248\pm25\degr$, in agreement with previous optical results \citep[e.g. $167\pm18\kmps$ from][adjusted for the GAIA distance]{Dougados2000}. Using other observed separations of knot A from DG Tau A  \citep[between epochs 1992.86 and 2010.15,][their Table 2]{Rodriguez2012DGTauA} and those observed here, we can produce the most accurate velocity for it to date. A least squares fit of the separations yields a velocity, in the plane of the sky, of $158\pm12\kmps$, corresponding to an ejection date of $1984.93\pm0.36$. This agrees well with the result from \citet{CohenBieging1986} whereby DG Tau A underwent a $\sim20$~percent increase in flux density and shift of the $5\GHz$ emission to the southwest (along the jet axis), between epochs 1983.90 and 1985.34. Adjusting for inclination, we calculate an absolute velocity for knot A of $258\pm23\kmps$.

As for the other radio sources, no clear detection of motion ($0.61\pm0\farcs54$) is detected towards knot C. DG Tau A, however, displays an apparent proper motion of $0.17\pm0\farcs05$, corresponding to a velocity of $25\pm7\kmps$, at a position angle of $88\pm14\degr$. We can compare this proper motion with that derived, over 30 years of radio observations, by \citet[][see their subsection 2.1]{Rodriguez2012DGTauA}. Using their results, we calculate that DG Tau A should, at epoch 2016.15, have moved (since 2012.22) $33\pm4\mas$ in right ascension and $-74\pm4\mas$ in declination, due to its motion through the TMC. Subtracting this from the apparent proper motion we have derived gives a velocity of $24\pm7\kmps$ at a position angle of $60\pm16\degr$, parallel with the jet's axis. This lends further support to the findings of \autoref{sec:variability} whereby DG Tau A has recently ejected a knot of emission towards the north east, since the radio emission's centroid should shift in the direction of any recent ejection.  

\begin{table*}
\begin{center}
\caption{Table of parameter values employed in the producing the model of the DG Tau A radio jet. All parameter names are taken from \citet{Reynolds1986}.}
\begin{tabularx}{\textwidth}{ccccZ}
\toprule
Parameter  & Description                              & Value               & Units             & Notes \\
\midrule
$\alpha$   & Flux density spectral index              & 0.4                 &                   & Measured value \\
$r_0$      & Launching radius                         & 0.212               & $\au$             & Dust sublimation radius$^\dagger$ \\
$\chi_0$   & Initial ionisation fraction              & 0.1                 &                   & Assumed from typical values in the literature \\
$q_{\rm T}$& Power law index for temperature          & 0                   &                   & No cooling or heating in the jet's stream\\
$q_{\rm v}$& Power law index for velocity             & 0                   &                   & No acceleration or deceleration after launch\\
$q_\upchi$ & Power law index for ionisation fraction  & 0                   &                   & No recombination or ionisation in the jet's stream\\
$\epsilon$ &  Power law index for jet width           & $+\nicefrac{7}{9}$  &                   & Inferred using \autoref{eq:epsilon} \\
$q_{\rm n}$      &  Power law index for number density      & $-\nicefrac{14}{9}$ &                   & Inferred using $q_{\rm n} = - q_{\rm v} - 2\epsilon\,^\ddagger$\\
$q_\uptau$ & Power law index for optical depth        & $-\nicefrac{7}{3}$  &                   & Inferred using $q_\uptau = \epsilon+2q_\upchi+2q_{\rm n}-1.35{q_{\rm T}}\,^\ddagger$\\
$w_0$      & Initial jet width                        & 0.152               & $\au$             & Inferred from observations using \autoref{eq:epsilonpowerlaw} \\
$n_0$      & Initial jet number density               & $2.6\times10^9$     & $\cmc$            & Inferred using $n_0 = n\left(r\right)\left(\tfrac{r_0}{r}\right)^{q_{\rm n}}\,^\ddagger$\\
$\dot{M}_\mathrm{jet}$ & Jet mass loss rate (per jet) & $3.7\times10^{-8}$  & $\Msol\,\yr^{-1}$ & Inferred using $\dot{M}_\mathrm{jet}=n_0 \pi \mu {w_0}^2 {v_0}\,^\ddagger$\\
\bottomrule
\multicolumn{5}{l}{$^\dagger$ \citet[adjusted for the Gaia distance]{Akeson2005}, $^\ddagger$ \citet{Reynolds1986}} \\
\label{tab:modelparams}
\end{tabularx}
\end{center}
\end{table*}

\begin{figure*}
	\includegraphics[width=5.329in]{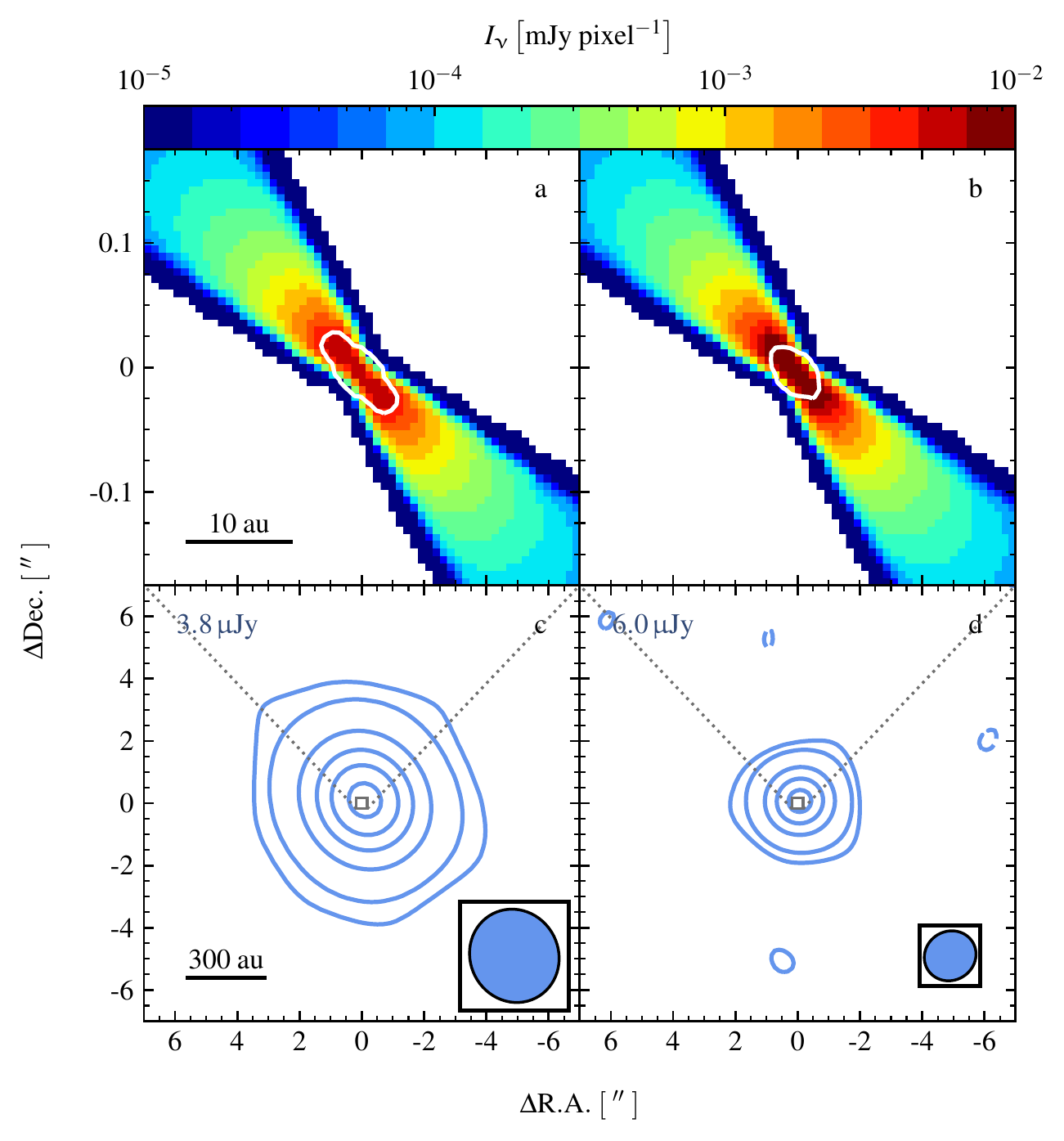}
	\caption{Plots of the intensity (colourscale) models used at $6$ and $10\GHz$ (panels a and b respectively) and the synthetic clean maps produced as a result of simulated observations towards those models (panels c and d respectively). In panels a and b, the cell size was set to $0.5\au$ and the surface whereby $\tau=1$ is indicated as a white contour. For panels c and d, the contour levels are set to $-3, 3, 10, 50, 100, 150$ and $200$ times the noise level, which is indicated in the top left corner of each panel. Restoring beams employed in the deconvolution are indicated in the bottom right of panels c and d, with restoring beam sizes of $3\farcs04 \times 2\farcs84$ at $\theta_\mathrm{PA} = 4\fdg\;\!\!8$ and $1\farcs69 \times 1\farcs57$ at $\theta_\mathrm{PA} = 54\fdg\;\!\!7$ respectively.}
	\label{fig:radiortmodel}
\end{figure*}

\subsection{Nature of the non-thermal emission}
\label{sec:nonthermal}
A work by \citet{Rivera2015} established the relative internal motion of DG Tau A within the TMC to be $-4.69\kmps$, $+1.02\kmps$ and $-2.32\kmps$ in the $u$, $v$ and $w$ directions respectively. In that work, the $(u,v,w)$ coordinate system was defined whereby $u$ is in the direction of the galactic centre, and $v$ and $w$ are parallel with the galactic longitude and latitude axes respectively. In the equatorial (J2000) coordinate system, this corresponds to an internal motion vector ($\mathbf{v_\mathbf{DGT|TMC}}$) of $1.65\kmps$ at a position angle of $150\fdg\;\!\!4$ in the plane of the sky (grey arrow in panel d of \autoref{fig:cplots2016}). Since their work adopted a distance of $150\pc$, we adjust this internal motion vector for the new Gaia distance and recalculate it to be $1.33\kmps$.

As shown in \autoref{tab:spix}, there are two knots of non-thermal emission (knots C and D) associated with DG Tau A's jet. Using previous GMRT results \citep{Ainsworth2014,Ainsworth2016}, in conjunction with the 2012 and 2016 VLA data, we derive $\alpha=-0.9\pm0.1$ and $-1.2\pm0.1$ for C and D respectively. For this calculation, only the 2012 VLA data (almost co-eval with the GMRT observations) was used for C due to its established variability (see \autoref{sec:variability}). \citet{Ainsworth2014} detect both knots in their GMRT data and suggest knot C to be a limb-brightened bow-shock from the jet. However, as demonstrated in \autoref{sec:propermotions}, no proper motion in the direction of the jet was seen and therefore this conclusion seems less likely. It is possible that both, or either, of the two non-thermal knots could be unrelated background objects. However, the probability of both being extragalactic in nature is $\sim10^{-6}$ \citep[based upon the previous calculation by][]{Ainsworth2014}.

With the calculated relative motions of DG Tau A within the TMC it is possible that, in DG Tau A's reference frame, dense cloud material (i.e.\ the density gradient alluded to in \autoref{sec:variability}) could move into the path of the jet. In turn this causes external, quasi-stationary shock sites at the working surfaces where the jet's `edges' make contact. This scenario would explain why no proper motions are found towards knot C (see \autoref{sec:propermotions}) and also why the non-thermal knots, C and D, are distinctly offset from the jet's outflow axis. \citet{Hartigan2005} observed a morphologically similar `deflection shock' towards the HH 47 jet in HST H$\alpha$ and $[\ion{S}{II}]$ images (their Figure 3). In that example, a stationary shock, resulting from the interaction of the HH 47 jet with ambient material, was offset from the jet's axis. Interestingly the linear morphology of that shock is similar to that of DG Tau A's knots C and D, where the major axes of these knots are similarly aligned in comparison to that of the jet, as in the case of HH 47 and its deflection shock. 

Should knots C and D be oblique shocks, using their deconvolved position angles (see \autoref{tab:results2016}) we calculate the angles between the jet's axis and the planes of the working surfaces at knots C and D to be $29\pm4\degr$ and $31\pm6\degr$ respectively (where $90\degr$ is the case of a head-on/perpendicular shock). We calculate this measured obliquity to decrease the effective speed of the shock by a factor of $\sim0.5$. Assuming the jet is impacting upon a static surface, the Mach number of the shock ($M_1=v_\perp / c_\mathrm{s}$, where we assume $c_\mathrm{s}=10\kmps$) still satisfies the limit for very strong shocks, whereby $M_1=14\pm2$. At astrophysical shock fronts, it is known that particles are accelerated to relativistic velocities by diffusive shock acceleration \citep[DSA;][]{Bell1978}. Assuming that jet axis and magnetic field direction are parallel \citep[i.e.\ as in the case of HH 80--81,][]{CarrascoGonzalez2010}, the obliquity of the shock affects the efficiency with which particles can be injected into the DSA mechanism by an order of magnitude \citep[estimated from Figure 6 of][]{Ellison1995}. Although potentially still a source of low-energy cosmic rays, this would detrimentally affect the cosmic ray production rates in this particular example, unless this stationary shock had been present for a long period of time.

\subsection{Modelling the radio jet}
\label{sec:modelling}

Throughout studies of thermal radio jets in the literature, the basic power-law models of \citet{Reynolds1986} are employed to understand their nature through radio observations. However many observations point to a more complex jet morphology with ejection variability, cross-sectional profiles in density, temperature and velocity to name but a few. Here we investigate if the simple power-law prescription can effectively model observations of thermal jets. Therefore, using radiative transfer calculations, we have produced synthetic images of DG Tau A's ionised jet in order to predict the distribution of ionised gas and its subsequent imaging with interferometers. For this, a Reynolds power-law model is used. To set up this model we need prior knowledge of the system's geometry, as well as intrinsic properties of the ionised gas. Measurement of the new emission discussed in \autoref{sec:variability} supplies these parameters. If we assume that the recent ejection/internal shock is completely ionised, separated from the jet-launching point by $352\pm60\au$ ($1.78\pm0\farcs29$ in the plane of the sky) and has a deconvolved diameter of $101\pm34\au$ (or $0.84\pm0\farcs28$), we can define the jet's width at any point along its axis. For this we employ the power-law defined in \autoref{eq:epsilonpowerlaw} using a calculated value for $\epsilon$ of $\nicefrac{7}{9}$, inferred from $\alpha\sim0.4$ for the thermal jet in conjunction with \autoref{eq:epsilon} \citep{Reynolds1986}. 

\begin{align}
w\left(r\right) &= w_0\left(\dfrac{r}{r_0}\right)^\epsilon\label{eq:epsilonpowerlaw}\\
\epsilon &= \dfrac{\alpha\left(2q_\upchi-2q_{\rm v}-1.35q_{\rm T}\right)-4q_\upchi+4q_{\rm v}+0.6q_{\rm T}-2.1}{3\alpha-3.9} \label{eq:epsilon}
\end{align}

\noindent where $w\left(r\right)$ is the width of the jet at a distance, $r$, along it's axis, $w_0$ is the width of the jet at the launching radius, $r_0$, and $q_\upchi$, $q_{\rm v}$ and $q_{\rm T}$ are the power-law exponents for ionisation fraction, velocity and temperature with $r$, respectively.

Using the derived electron density for the ejected lobe of $n_{\rm e}=(2.0\pm1.1)\times10^4\,\mathrm{cm^{-3}}$ as a measure of the total density (i.e.\ $\chi_i=1$), a basic model of the jet can be computed in the form of a spatial grid of densities, pressures and temperatures. It is assumed that there is no recombination, acceleration or cooling for this model, the parameters of which are listed in \autoref{tab:modelparams} with reasoning for each assumption given in the last column. 

Following their computation, the density, pressure and temperature grids (a cell size of $0.5\au$ was employed) for DG Tau A were used as the input into \texttt{RadioRT}, a radio continuum and recombination line radiative transfer code \citep{Dougherty2003,Steggles2017}, at simulated frequencies of $6$ and $10\GHz$. Products of the code were images of emission measure (units of $\mathrm{pc\,\,cm^{-6}}$), intensity (units of $\mathrm{mJy\,\,pixel^{-1}}$) and optical depth (dimensionless), of which the $6$ and $10\GHz$ intensity images are shown in panels a and b of \autoref{fig:radiortmodel} respectively. For reference, the surface where the optical depth is unity is shown as a white contour in each plot, highlighting the compact ($79\mas$ and $51\mas$, or $9.5\au$ and $6.2\au$, along its major axis at $6$ and $10\GHz$ respectively) nature of the optically thick emission.

From the intensity model, we measure a spectral index of $\alpha_\mathrm{model}=0.48\pm0.01$. Errors are propagated normally for $\alpha_\mathrm{model}$, with the errors on derived fluxes being $\sigma_{S_\upnu}=\nicefrac{\Sigma I_\upnu}{\sqrt{N}}$, where $\Sigma I_\upnu$ is the sum of pixel intensities and $N$ is the number of pixels. While seemingly at odds to that predicted by \citet{Reynolds1986} (see \autoref{tab:modelparams}), this calculation of the spectral index sums pixel intensities from both optically thick, and thin, parts of the jet (i.e.\ all pixels). In actual fact, the spectral index predicted by \citet{Reynolds1986} comments only on the spectral index of the optically thick regions of the emission, $\alpha_{\rm op}$. Recalculating the spectral index whilst excluding optically thin pixels gives $\alpha_{\rm op}=0.55\pm0.45$, with the large error in $\alpha_{\rm op}$ being the product of the limited number of pixels (i.e.\ resolution) over which the intensity was summed. Further investigation was performed by increasing the pixel resolution of our models ($\Delta x=0.1\au$), and repeating the same calculation. In this case we derive a spectral index, $\alpha_{\rm op}$, of $0.41\pm0.05$, in agreement with \citet{Reynolds1986}.

Intensity images from \texttt{RadioRT} formed the sky-model for subsequent synthetic observations using \textsc{casa}'s \texttt{simobserve} task, for which all instrumental, environmental (i.e.\ noise) and observational parameters were set to match those of the relevant 2016 observations. After production of the synthetic visibility datasets, standard imaging and deconvolution was performed using a uniform robustness of $R=-2$ to maximise resolution. Resulting synthetic images are shown in panels c and d of \autoref{fig:radiortmodel} for both frequencies, while measured flux densities and dimensions for the emission are tabulated in \autoref{tab:radiortmodel}. We calculate a value for $\alpha$ of $0.49\pm0.03$, however no physical dimensions could be deconvolved due to the highly compact nature of the emission (roughly the $\tau_\upnu=1$ surfaces shown in panels a and b of \autoref{fig:radiortmodel}). Our derived spectral index is higher than $\alpha_{\rm op}$ since both optically thick, and thin, emission is contained within the synthetic observations' beams. For future jet studies this is important when utilising the models of \citet{Reynolds1986}. We believe that the angular scale of both the thick, \textit{and thin}, emission, in relation to the synthesised beam of the observations, must be taken into account in order to properly interpret the jet's physical conditions on the basis of $\alpha$.

Comparing flux densities of the model to those measured from the 2016 data, it seems that the model slightly overestimates DG Tau A's observed flux density by $\sim10$~percent. This over-estimation is may be due to our assumption of full ionisation in the shock, used to calculate the initial density, $n_0$, of the jet. However, the observed physical dimensions are not reproduced by the model, with the jet remaining unresolved in the synthetic observations. To account for the observed, extended emission, either opening angles are much smaller, or that re-ionisation of jet material at working surfaces is present along the jet's stream. In light of previous optical imagery \citep[i.e. Figures 2 and 3 of][]{Dougados2000}, modelling \citep{Raga2001} and the results of \autoref{sec:variability}, we believe the second possibility to be much more likely, on the basis of the extent of the emission (opening angles from $\sim11$ to $\sim33\degr$), variability of the DG Tau A jet, as well as potential precession of the outflow axis (which should lead to more external working surfaces).

\begin{table}
\begin{center}
\caption{Table of derived values for flux density and deconvolved dimensions for the imaged model of the DG Tau A radio jet.}
\begin{tabularx}{\columnwidth}{ZZZZZ}
\toprule
$\nu$ & $S_\upnu$ & $\theta_\mathrm{maj}$ & $\theta_\mathrm{min}$ & $\theta_\mathrm{PA}$\\
$\left[\mathrm{GHz}\right]$ & $\left[\!\!\uJy\right]$ & $\left[\!\!\mas\right]$ & $\left[\!\!\mas\right]$ & $\left[\degr\right]$ \\
\midrule
6 & $830\pm7$ & - & - & - \\
10 & $1066\pm11$ & $<510$ & $<140$ & - \\
\bottomrule
\label{tab:radiortmodel}
\end{tabularx}
\end{center}
\end{table}

\section{Conclusions}
\label{sec:conclusions}
In this work we have performed follow-up 6/10$\GHz$, VLA observations (epoch 2016.10) of the jet associated with the YSO DG Tau A, in order to examine the nature of its radio emission.

In conjunction with the reduction of a previous epoch's (2012.22) data we have been able to spatially resolve the radio variability towards DG Tau A's jet and associated working surfaces on scales of $\geq200\au$ and confirm, or establish, proper motions of all radio components. From this analysis we are able to conclude the following:
\begin{enumerate}
\item For the first time at $\cm$ wavelengths, we detect the non-thermal `counterjet' (our knot D) previously seen in GMRT observations. As with the other non-thermal knot (C), it is offset from the jet's axis towards the SE.
\item No polarization is seen towards DG Tau A's radio jet and associated knots A, C and D with $3\upsigma$ upper limits in linear polarization of $<1.3$, $<50.8$, $<18.2$ and $<51.5$~percent respectively.
\item Although DG Tau A's overall flux density has decreased over the last 4 years, we observe an increase in flux density along the receding jet's axis. We conclude that DG Tau A's receding jet has undergone a variable ejection event, which has not been seen previously, with an average, ionised mass loss rate of $\left(3.7\pm1.0\right)\times10^{-8}\Msol\yr^{-1}$. This behaviour is not contemporaneously seen in the approaching jet, showing that time variable mass loss is an asymmetric process.
\item Over a period of $\sim4\yr$, and in agreement with previous observations, we observe a proper motion of $1.6\pm0.6$ arcsec, at a position angle of $248\pm25\degr$, in radio knot A. In conjunction with previous data, we consequently derive an absolute velocity in the approaching jet of $258\pm23\kmps$.
\item No proper motions are observed towards radio knot C, which was previously thought to be the limb of an optical bow-shock. In conjunction with the offset of knot C from the jet's axis, we instead conclude this to be a static shock upon a working surface produced by the impingement of jet material upon a density gradient present to the SE. This is supported by a spatially-resolved increase of flux density, over the last 4 years.
\item From modelling of the radio jet, in order to adequately explain the physical extent of the emission, shocks along the jet surface, leading to re-ionisation of the material, must be present.
\item Future radio observations of jets must take into account the scales of both the optically thick, and thin, emission in relation to the synthesised beam of the observations, in order to accurately interpret jet physical conditions from spectral index values.
\end{enumerate}

Further, sensitive, radio observations of DG Tau A and its jet in the future will be required to establish the working surface nature of knot C, as well as refine the velocity estimates of the recent ejection of jet material in the receding jet. Pushing down the limits on the degree of linear polarization, with even more sensitive observations, will be key in establishing how this low mass radio jet is collimated on larger scales, as in previous examples of YSOs.
\section*{Acknowledgements}
SJDP and TPR would like to acknowledge support from the advanced grant H2020--ERC--2016--ADG--74302 from the European Research Council (ERC) under the European Union's Horizon 2020 Research and Innovation programme.

This work has made use of data from the European Space Agency (ESA) mission {\it Gaia} (\url{https://www.cosmos.esa.int/gaia}), processed by the {\it Gaia} Data Processing and Analysis Consortium (DPAC, \url{https://www.cosmos.esa.int/web/gaia/dpac/consortium}). Funding for the DPAC has been provided by national institutions, in particular the institutions participating in the {\it Gaia} Multilateral Agreement.

The National Radio Astronomy Observatory is a facility of the National Science Foundation operated under cooperative agreement by Associated Universities, Inc.

Throughout this work we also made use of \texttt{astropy}, a community-developed core python package for astronomy \citep[version 3.0.1,][]{Astropy2013}, and \texttt{uncertainties}, a python package for calculations with uncertainties (version 3.0.1) developed by Eric O. Lebigot, for plotting and error propagation purposes respectively.


\bibliographystyle{mnras}
\bibliography{Biblio} 


\bsp	
\label{lastpage}
\end{document}